\documentstyle[11pt,fleqn]{article}
\def\beq{\begin{equation}}
\def\eeq{\end{equation}}

\newtheorem{theorem}{Theorem}[section]
\newtheorem{lemma}[theorem]{Lemma}
\newtheorem{proposition}[theorem]{Proposition}
\newtheorem{corollary}[theorem]{Corollary}

\newenvironment{proof}{\begin{trivlist}\item[]{\em Proof.\/\ }}%
                      {\hfill$\Box$\end{trivlist}}
                      {\hfill$\Box$\end{trivlist}}
                      {\hfill$\Box$\end{trivlist}}
\newenvironment{lemma*}{\begin{trivlist}\item[]{\bf Lemma.\ }\em}%
                      {\end{trivlist}}
\newenvironment{theorem*}{\begin{trivlist}\item[]{\bf Theorem.\ }\em}%
                      {\end{trivlist}}
\newenvironment{corollary*}{\begin{trivlist}\item[]{\bf Corollary.\ }\em}%
                      {\end{trivlist}}

\def\remark{\refstepcounter{theorem}\par\noindent{\bf Remark \thetheorem\ }}
\def\remarks{\refstepcounter{theorem}\par\noindent{\bf Remarks \thetheorem\ }}
\def\nextrem{\refstepcounter{theorem}\par\noindent{\bf \thetheorem\ \ \ }}

\def\calH{{\cal H}}

\def\Bbb{\bf}
\newcommand{\RR}{{\Bbb R}} \newcommand{\ZZ}{{\Bbb Z}} 
 \newcommand{\CC}{{\Bbb C}} 
\newcommand{\QQ}{{\Bbb Q}} \newcommand{\TT}{{\Bbb T}}

\def\SC{\mbox{\scriptsize sc}}

\def\UH#1{U$#1$H}
\def\Lone{\mbox{\rm L}^1}
\def\Ltwo{\mbox{\rm L}^2}
\def\ae{a.e.\ }

\def\calH{{\cal H}}

\def\char#1{e^{-2\pi i {#1}}}
\def\cchar#1{e^{2\pi i {#1}}}
\def\gms{{\gamma_{\mbox{\scriptsize ms}}}}
\def\gmsl{{\gms(\lambda)}}
\def\gom{{\gamma(\omega)}}
\def\sno{{S_n^\omega}}
\def\snol{{\sno(\lambda)}}
\begin{document}

\title{On scaling in relation to singular spectra}

\author{A. Hof\thanks{
        Division of Physics, Mathematics and
        Astronomy, California Institute of Technology 253-37,
        Pasadena, CA 91125. E-mail: hof@cco.caltech.edu\newline
        To appear in {\em Communications in Mathematical Physics.\/}
        }
        }

\date{}
\maketitle

\begin{abstract}
This paper relates uniform $\alpha$-H\"older continuity,
or $\alpha$-dimensionality, of spectral measures in an arbitrary interval
to the Fourier transform of the measure.
This is used to show that scaling exponents of exponential sums
obtained from time series give local upper bounds on the degree
of H\"older continuity of the power spectrum of the series.
The results have applications to generalized random walk, numerical detection
of singular continuous spectra and to the energy growth in driven oscillators.
\end{abstract}

\section{Introduction}

An interesting method to numerically detect singular continuous 
spectum has been proposed by Aubry, Goldr\`eche and 
Luck \cite{Aub+87a,Aub+88}. 
In essence, their method amounts to the following.
Let $\{a_k\}_{k=0}^\infty \subset \CC$ and  
$S_n(q):= n^{-1} \bigl| \sum_{k=0}^{n-1} a_k \char{kq}\bigr|^2$.
Suppose that $S_n(q)\,dq$
converges weakly as $n\to\infty$ to a positive Borel measure
$\mu$ on $\TT=\RR/\ZZ$. 
Then $\mu$ should be singular at values of $q$ for which 
$S_n(q)$ `scales' like $n^{\beta(q)}$ for $0 < \beta(q) <1$.
In a previous paper \cite{Hof96a} we proved, without justifying
any scaling argument, that the particular
model Aubry et al.\ considered does indeed have purely singular 
continuous spectrum for generic parameters (see Section~5.1
for details).

The method of Aubry et al.\ has been used
to argue that sequences obtained from aperiodically driven 
quantum systems \cite{Luc+88,Oli95} and a time series obtained 
from a dynamical system  \cite{PiFe94,Pik+95} have singular
continuous power spectrum.
Moreover, the growth of sums of the form $S_N(q)$ has been shown
to detemine the growth of the energy in driven classical and
quantum-mechanical oscillators \cite{Bun+91,Com92}.
Finally, Luck \cite{Luc89} has perturbatively related the 
exponents $\beta(q)$ to 
the speed with which gaps close in the spectrum of a discrete 
Schr\"odinger operator with potential $\lambda a_n$ as 
$\lambda\to 0$.

It is desirable, therefore, that the mathematical meaning of the
exponents $\beta(q)$ and their relation to the singularity of the
measure $\mu$ is clarified.
That is the aim of this paper.
It will show that: (i)
if $\{a_n\}$ is  a time series of an
ergodic dynamical system, then $\beta(q)$ exist \ae with 
respect to the ergodic measure when interpreted as
critical exponent for a limsup, and (ii) $1-\beta(q)$ is an upper 
bound on the degree of H\"older continuity of $\mu$ in a 
neighborhood of $q$ (see Corollary~\ref{cor2}).
However, scaling exponents $\beta(q)\in (0, 1)$ need not 
imply that $\mu$ has a singular continuous part (see Remarks~\ref{beta01}
and \ref{scpart}).

H\"older continuity here means uniform H\"older continuity.
A positive Borel Measure on $\RR$ is uniformly $\alpha$-H\"older continuous
--- \UH\alpha, for short --- in an interval $I$
if there is a constant $C$ such that $\mu(I_0) \leq C|I_0|^\alpha$
for all intervals $I_0 \subset I$ with $|I_0|<1$, where $|I_0|$ denotes the
length of $I_0$.
It is globally uniformly $\alpha$-H\"older continuous 
if it is  \UH\alpha\ on its support.
Uniform $\alpha$-H\"older continuity is often referred to as 
uniform $\alpha$-dimensionality (e.g., in \cite{Str90}).
A measure that is \UH\alpha\ gives zero weight to sets
that have measure zero for $\alpha$-dimensional Hausdorff measure
(cf.\ Section~3.3 in \cite{Rogers}).

Section~2 of this paper relates a double integral over
the Fourier transform $\hat\mu(t):= \int \char{tx} \, d\mu(x)$ 
of a finite positive Borel measure---the convolution of the measure
with the Fej\'er kernel---
to the degree of H\"older continuity in arbitrary
intervals.
This should be contrasted to results that relate global uniform
H\"older continuity to the averaged decay of $\hat\mu$.
Strichartz \cite{Str90} has shown that if $\mu$ is globally \UH\alpha\ then
there is a constant $C$ such that 
$T^{-1}\int_0^T |\hat\mu(t)|^2 \, dt < CT^{-\alpha}$.
A partial, but optimal, converse has been obtained by Last
\cite{Las96}: if there exists a constant $C$ such that 
$T^{-1}\int_0^T |\hat\mu(t)|^2\, dt < CT^{-\alpha}$ for all $T>0$ then
$\mu$ is globally \UH{\frac\alpha 2}.
Of course a measure may very well have a degree of 
uniform H\"older continuity $\alpha$ that varies over its support. 
Then it is the most singular part (i.e., the part with the smallest 
$\alpha$) that matters for the results just mentioned.
Global uniform H\"older continuity of spectral measures of Schr\"odinger 
operators has received attention because it gives lower bounds 
on the spread of wave functions \cite{Gua89,Com93}, see also 
\cite{Gua93,GuMa94,Las96,JiLa96,Rio+96,Bar+96}.

The remaining sections apply the result of Section~2
to spectral measures of dynamical  systems.
Section~3 relates the degree of H\"older continuity
of a spectral measure  near zero to critical exponents of `generalized
random walks'.
The generalization to `twisted generalized random walks'
in Section~4 explains the meaning of the scaling exponents
$\beta(q)$.
Finally, Section~5 discusses the application to the model
considered by Aubry et al., and to the energy growth 
of driven oscillators.

\section{Uniform H\"older continuity in an interval} 

This section explains how the Fourier transform of a positive bounded
Borel measures can be related to uniform H\"older continuity in
arbitrary intervals. 
In view of the applications that follow, the result will be stated
and proved for spectral measures of a unitary operator.
A series of remarks discusses some generalizations and the
relation to the decomposition of the measure into its
discrete, singular continuous and absolutely continuous 
parts.

Let $U$ be a unitary operator on a separable Hilbert space $\calH$.
Every $\psi\in\calH$ defines a spectral measure $\mu_\psi$ on
the circle $\TT$ by $\int_\TT  \cchar{nx} \, d\mu_\psi = (U^n\psi, \psi)$.
For $\lambda\in\TT$ let $U_\lambda:=\char{\lambda}U$ and
$G_n(\lambda):= \frac1n \bigl\| \sum_{k=0}^{n-1} U^k_\lambda \psi \bigr\|^2$.
Then $\lim_{n\to\infty} G_n(\lambda)\,d\lambda= \mu_\psi$ weakly (as can
e.g.\ be seen by integrating $\char{m\lambda}$ with respect to 
$G_n(\lambda)\,d\lambda$).
The following theorem says that $G_n(\lambda)$
scales like $n^\beta$ as $n\to\infty$ if and only if
$\mu_\psi$ is \UH{(1-\beta)} in a neighborhood of $\lambda$.

\begin{theorem}\label{thm1}
Let $\Delta$ be an open interval and $0\leq\beta\leq 1$. 
The following two statements are equivalent:\\
i) There exists a constant $C>0$ such that 
    $\limsup_{n\to\infty} n^{-\beta}G_n(\lambda) \leq C<\infty$
   uniformly in $\lambda\in\Delta$;\\
ii) $\mu_\psi$ is \UH{(1-\beta)} in $\Delta$.
\end{theorem}
\begin{proof}
Direct computation shows that 
$G_n(\lambda) = \int_\TT K_{n-1} (\lambda-x)\, d\mu_\psi(x)$, where
$K_{n-1}(y):= {\frac 1 n} \bigl( {{\sin n\pi y}\over{\sin\pi y}} \bigr)^2$
is the Fej\'er kernel.
Note that $K_n(0)=n$ for all $n$ and that 
$\lim_{n\to\infty}K_n(y)=0$ uniformly on every closed interval not
containing 0.

{\em i)\/} implies {\em ii).\/} By  {\em i)\/} there exists for
every $\epsilon>0$  an $N>0$ such that 
$G_n(\lambda) \leq (C+\epsilon)n^\beta$ for all $n>N$ and all 
$\lambda\in\Delta$.
Let $I\subset \Delta$ be an interval of length $|I|<1/N$.
Let $n>N$ be such that $\frac 1n < |I| \leq \frac 1{n-1}$.
Then $I$ can be covered by two adjacent intervals 
$I_j:= [\lambda_j - \frac 1{2n}, \lambda_j + \frac1{2n}]$ of
length $1/n$.
Note that $\frac{\pi^2}{4n}K_{n-1}(\lambda_j-x)\geq1$
if $x\in I_j$.
Therefore,
\begin{eqnarray*}
 \mu_\psi(I) 
  &\leq& \sum_{j=1}^2 \int_{I_j} \frac{\pi^2}{4n} 
              K_{n-1}(\lambda_j-x) \, d\mu_\psi(x)\\
  &\leq& \frac{\pi^2}{4n} \bigl\{ G_n(\lambda_1) + G_n(\lambda_2) \bigr\} \\
  &\leq& \frac{\pi^2}4 (C+\epsilon) |I|^{1-\beta}.
\end{eqnarray*}

{\em ii)\/} implies {\em i)\/}.
Let $\lambda\in\Delta$ and choose $0<a< 1/2$ such that 
$[\lambda-a, \lambda+a] \subset \Delta$.
It suffices to show that 
\[
 n^{-\beta} \int_{\lambda-a}^{\lambda+a} K_{n-1}(\lambda-x)\, 
   d\mu_\psi(x) \leq C < \infty
\]
for a constant $C$ independent of $\lambda$ and $n$.
Since $\mu_\psi$ is \UH{(1-\beta)} in $\Delta$ there is a constants
$D, N$ such that $\mu_\psi(I) \leq Dn^{\beta-1}$ for every interval 
$I$ of length $1/n$ with $n>N$.
Now $K_{n-1}(\lambda-x)\leq n$ if $|\lambda-x|< 1/n$ and 
$K_{n-1} (\lambda-x) \leq 4n/k^2\pi^2$ if 
$k/n \leq |\lambda-x| \leq (k+1)/n$ since $\sin \pi y \geq y/2$
for $0\leq y\leq 1/2$.
Therefore, denoting the integer part of $an$ by $[an]$,
\[
 \int_{\lambda-a}^{\lambda+a} K_{n-1}(\lambda-x)\, d\mu_\psi(x)
  \leq 2Dn^\beta + 2\sum_{k=1}^{[na]} \frac4{k^2\pi^2} Dn^\beta
    \leq Cn^\beta,
\]
for some constant $C$ independent of $\lambda$ and $n$.
\end{proof}

\remarks
Theorem~\ref{thm1} holds in greater generality than stated.
Let $\mu$ be a finite positive Borel measure on $\RR$ 
with Fourier transform $\hat\mu(t)=\int \char{tx} \, d\mu(x)$.
For $T>0$, define
$G_T(\lambda):= T^{-1} \Bigl| \int_0^T \int_0^T \cchar{(t-s)\lambda}
     \hat\mu(t-s) \, ds\,dt\Bigr|^2$.
Again, $G_T(\lambda)=\int K_T(\lambda-x) \,d\mu(x)$, where 
\[
   K_T(y) = \int^T_{-T} (1- {\frac {|u|}T})\cchar{yu}\, du 
        = \frac 1T \Bigl({ sin \pi T y \over \pi y} \Bigr)^2
\]
is the Fej\'er kernel on $\RR$.
Now Theorem~\ref{thm1} holds for $\mu$ and this function $G_T$.
The proof is the same; that of `{\em i)\/} implies {\em ii)\/}'
simplifies in that one can take $T=|I|^{-1}$.
\nextrem\label{rem2.3}
In particular, if $\mu=\mu_\psi$ is a spectral measure of 
a strongly continuous group $\{U_t\}_{t\in\RR}$ of unitary
operators on a separable Hilbert $\calH$ 
(i.e.\ $\int_\RR \cchar{tx} \,d\mu_\psi(x) = (U_t\psi, \psi)$ 
for a $\psi\in\calH$) then Theorem~\ref{thm1} holds for $\mu_\psi$
and 
$G_T(\lambda) := T^{-1} \Bigl\| \int_0^T 
    U_{\lambda, t} \psi\, dt\Bigr\|^2$,
where $U_{\lambda, t}:= \char{\lambda t} U_t$. 

\nextrem
Theorem~\ref{thm1} also generalizes to higher dimensions.
This is of interest for actions of $\ZZ^d$ on a
probability space $(\Omega, \nu)$ that leave $\nu$
invariant.

Let $U^k$, $k\in\ZZ^d$, be the unitary action of $\ZZ^d$
on $\Ltwo(\Omega,\nu)$ defined by $U^k f := f\circ T^k$,
where $T^k$ is the action of $\ZZ^d$ on $(\Omega, \nu)$.
For $\lambda\in\TT^d$, let $U_\lambda^k:=
\char{\langle \lambda, k \rangle} U^k$, where
$\langle\,\cdot\, , \,\cdot\, \rangle$ denotes the 
Euclidian inner product in $\RR^d$.
Let $C_n:= \bigl\{ m\in \ZZ^d \bigm| 0\leq m_i <n\bigr\}$
and $G_n(\lambda):= n^{-d} \bigl\| \sum_{k\in C_n}
U^k_\lambda \psi \bigr\|^2$. 
The Fej\'er kernel now is 
$K_n(y)= n^{-d} \prod_{i=1}^d \bigl(\sin n\pi y_i / \sin \pi y_i \bigr)^2$.
The theorem holds as before with $0\leq\beta\leq d$, 
$\Delta$ a cube, and \UH{(d-\beta)} instead of 
\UH{(1-\beta)} in statement {\em ii).\/}

\nextrem\label{beta01}
The values of $\lim_{n\to\infty} n^{-\beta} G_n(\lambda)$ 
for $\beta=1$ and $\beta=0$ determine the discrete part
and the absolutely continuous part of $\mu_\psi$, respectively.
Since the functions $n^{-1}K_n$ tend to zero uniformly outside 
any neighborhood of zero, and $n^{-1}K_n(0)=1$, one
has $\mu_\psi(\{\lambda\})=\lim_{n\to\infty} n^{-1} G_n (\lambda)$
for all $\lambda\in \TT$ (e.g., p.~42 in \cite{Katznelson}).
Thus $\beta=1$ determines the discrete part of $\mu_\psi$.
To see that $\beta=0$ determines the absolutely
continuous part of $\mu_\psi$, write $\mu_\psi = f \sigma + \rho$, 
where $\sigma$ denotes Lebesgue measure, $f\in \Lone(\sigma)$
and $\rho\perp\sigma$. 
Then $G_n(\lambda)\to f(\lambda)$ at $\sigma$-\ae $\lambda$,
see e.g.\ Theorem~III.8.1 in Volume~1 of \cite{Zygmund}.
If $f$ can be chosen to be continuous in a neighborhood
of $\lambda$ and $\rho$ is zero on that neighborhood then
$\lim_{n\to\infty} G_n(\lambda) = f(\lambda)$.

\nextrem\label{scpart}
A measure $\mu$ that is \UH\alpha\ for some $0<\alpha<1$
need not have a singular continuous part.
For instance, let $\{c_i\}$ be a positive sequence with
$\sum c_i <\infty$, $x_i \in\TT$, $0<\beta< 1$ and let
$\mu:= \sum c_i |x- x_i |^{-\beta}$.
Then $\mu$ is absolutely continuous and \UH{(1-\beta)}.
Still, the behavior of $n^{-\beta}G_n(\lambda)$ for 
$\beta=0$ and 1 can always be used in principle 
to determine whether $\mu_\psi$ has a singular 
continuous part $\mu_{\psi,\, \SC}$, because 
\[
  \mu_{\psi,\, \SC} (\TT) = \|\psi\|^2 - \int f\, d\sigma
          - \sum \lim_{n\to\infty} n^{-1} G_n(\lambda),
\]
where, as in the previous remark, $f=\lim_{n\to\infty} G_n$.

\nextrem\label{crithol}
The critical H\"older exponent $\alpha_\mu(x)$ of $\mu$ at $x$
is defined by
\[
  \limsup_{r\to 0} \mu([x-r, x+r]) r^{-\alpha} =
   \left\{
    \begin{array}{ll}
      0  & \mbox{\rm if $\alpha < \alpha_\mu(x)$} \\
      \infty & \mbox{\rm if $ \alpha > \alpha_\mu(x)$}
    \end{array}
   \right.
\]
Note that $\alpha_\mu(x)$ may be larger then $1$ if $\mu$ has
very little mass near $x$. 
If $-1\leq\beta\leq 1$ and $\limsup n^{-\beta} G_n(\lambda) <\infty$
then $1-\beta \leq \alpha_\mu(\lambda)$.
This follows from $\mu([\lambda-\frac1 {2n},\lambda+ \frac 1{2n}])
\leq \pi^2/4n \int_{|\lambda-x|\leq 1/2n} K_{n-1}(\lambda-x)\,d\mu(x)$.

\section{Generalized random walk}

Let $\Omega$ be a compact metric space with its Borel $\sigma$-algebra,
$T\colon\; \Omega\to\Omega$ a measurable invertible map and $\nu$ 
a $T$-invariant ergodic probability measure on $\Omega$.
Each $\psi\in\Ltwo(\Omega,\nu)$ gives rise to a 
so-called generalized random walk (GRW)
\beq\label{e3.1}
  S^\omega_n := \sum_{k=0}^{n-1} \psi (T^k \omega).
\eeq
Simple random walk on $\ZZ^2$ is the GRW 
defined by $\Omega=\{0,1,2,3\}^\ZZ$, $T$ the left shift on
$\Omega$, $\nu=\prod_{j\in\ZZ}[(\delta_{\omega_j, 0}+\delta_{\omega_j, 1} +
\delta_{\omega_j, 2}+\delta_{\omega_j, 3})/4]$ and 
$\psi(\omega)=\psi(\omega_j)=1$, $i$, $-1$, $-i$ if $\omega_j=0$, 1, 2,
and 3, respectively.
Generalized random walks, and especially their recurrence
properties, have been considered in e.g.\ 
\cite{AaKe82,Ber81,Dek82,WeWe92,Dek93,Dek95}.
The aim of this section is to relate the H\"older continuity at 0
of the spectral measure $\mu_\psi$ of $\psi$
to the speed with which the walk wanders off to 
infinity, as expressed by the critical exponents of the
mean squared displacement and of $|S^\omega_n|^2$.

Recall that the critical exponent $\alpha_c$ of a  sequence 
$N_n\in\CC$ is defined by 
\beq\label{e3.2}
  \limsup_{n\to\infty} n^{-\alpha} |N_n| =
   \left\{
    \begin{array}{ll}
      \infty & \alpha < \alpha_c \\
      0      & \alpha > \alpha_c
    \end{array}
   \right.
\eeq
The critical exponent of $|S_n^\omega|^2$ is denoted by $\gamma(\omega)$,
that of the mean squared displacement 
\[
  \bigl\langle |S^\omega_n|^2 \bigr\rangle 
   := \int |S_n^\omega|^2 \, d\nu(\omega) 
    = \Bigl\| \sum_{k=0}^{n-1} U^k \psi \Bigr\|^2_{\mbox{\scriptsize L$^2$}}
\]
by $\gms$; here $U\psi := \psi\circ T$ is the unitary operator
implementing the dynamics in $\Ltwo(\Omega,\nu)$.
Note that $0\leq \gms \leq2$.

\begin{lemma}\label{lem1}
There exists a $\gamma\in[0,\gms]$ such that 
$\gamma=\gom$ for $\nu$-\ae $\omega\in\Omega$.
\end{lemma}
\begin{proof}
For $\alpha>0$ and $I\subset[0,\infty)$ an interval, let
$B_{\alpha,I}:= \bigl\{ \omega\in\Omega \bigm| 
       \limsup_{n\to\infty} n^{-\alpha}|S_n^\omega|^2\in I \bigr\}$.
This set is measurable  since 
$B_{\alpha, I} = \bigcap_{K=0}^\infty \bigcup_{L=K}^\infty
 \bigl\{ \omega\in\Omega \bigm| 
   n^{-\alpha}|S_n^\omega|^2\in I \mbox{ for an } n\in[K, \ldots,L]\bigr\}$.
It is also invariant.
Hence the sets $B_{\alpha,0}:=\bigcap_{m=1}^\infty B_{\alpha, [0,1/m]}$
and $B_{\alpha,\infty}:= \bigcap_{m=1}^\infty B_{\alpha,[m,\infty)}$
are measurable and invariant.
Thus $\Omega$ is the disjoint union of the measurable and invariant
sets $B_{\alpha,0}$, $B_{\alpha, (0,\infty)}$ and $B_{\alpha,\infty}$.
By ergodicity, one of these three sets has measure one.
Since there is at most one value of $\alpha$ for which 
$\nu(B_{\alpha,(0,\infty)})=1$ there is a $\gamma$ such that 
$\nu(B_{\alpha,\infty})=1$ if $\alpha<\gamma$ and $\nu(B_{\alpha,0})=1$
if $\alpha>\gamma$.
This shows that $\gom = \gamma$ for $\nu$-\ae $\omega\in\Omega$.

If $\limsup_{n\to\infty} n^{-\beta} \bigl\langle |S^\omega_n|^2 \bigr\rangle 
=0$, then there is a subsequence $n_k$ such that 
$\lim_{k\to\infty}n_k^{-\beta} |S^\omega_{n_k}|^2 =0$ for $\nu$-\ae $\omega$.
This means that $\nu(B_{\beta,0})=1$.
Hence $\beta>\gms$ implies $\beta>\gamma$, and $\gamma\leq\gms$.
\end{proof}

It is now clear that the degree of H\"older continuity of $\mu_\psi$ 
at zero gives an upperbound on $\gms$, and that,
conversely, $\gms$ and $\gamma$ provide upperbounds on the 
degree of H\"older continuity of $\mu_\psi$ near zero.
This follows from the fact that, in the notation of 
Theorem~\ref{thm1}, $G_n(0)= n^{-1} \bigl\langle |S^\omega_n|^2 \bigr\rangle$.
If $\mu_\psi$ is \UH{\beta} on a neighborhood of 0, then 
$\gms\leq2-\beta$.
Conversely, if $\gms>1$ (or $\gamma>1$), then  
$\mu_\psi$ is not \UH{(2-\beta)} on any neighborhood of 0, for 
any $\beta\in(1,\gms)$ ($\beta\in(1,\gamma)$).
Thus superdiffusive behavior of the GRW requires that $\mu_\psi$
is not \UH1 on any neighborhood of 0.
On the other hand, the GRW can only be subdiffusive ($\gms<1$) if $\mu_\psi$
has very little weight near 0.
Indeed, by Remark~\ref{crithol} this requires $\alpha_{\mu_\psi}(0)>1$.
Theorem~1 in \cite{Lia87} gives that 
$\bigl\langle |S^\omega_n|^2 \bigr\rangle$ is bounded in $n$
if and only if $\int (\sin \pi \lambda)^{-2}\, d\mu_\psi(\lambda) <\infty$.

\medskip
\remarks\label{rudshap}
Dekking \cite{Dek95} shows that $\gms\leq 1$ for the `Rudin-Shapiro walk'. 
This is the GRW in which $\Omega\subset\{0,1,2,3\}^\ZZ$ is the substitution 
dynamical system  (e.g., \cite{Queffelec}) arising from the primitive 
substitution $0\to02$, $1\to32$, $2\to01$, $3\to31$ with 
$\psi(\omega)=\psi(\omega_0)= 1$, $-i$, $i$, $-1$ if
$\omega_0=0$, 1, 2, 3, respectively.
The spectral measure $\mu_\psi$ 
can be computed from Proposition~VII.5 and Example VIII.2.2 in
\cite{Queffelec}  as Lebesgue measure, so $\gms=1$, as conjectured by Dekking,
and the diffusion coefficient is 1.
(In Proposition~VII.5 of \cite{Queffelec} one should read 
$\overline{\tau(\beta)}$ for $\tau(\beta)$; this $\tau$ is 
our $\psi$.
The symbols $\{0,1,2,3\}$ in \cite{Queffelec} correspond to the
symbols $\{a,d,b,c\}$ in \cite{Dek95}.)

\nextrem\label{diff}
By Remark~\ref{beta01}, if $\mu_\psi$ is absolutely continuous on a 
neighborhood 0, with 
a density that can be chosen to be a continuous function $g$,
then $\gms=1$ and the diffusion coefficient of the GRW is $g(0)$.

\nextrem
Dumont and Thomas \cite{DuTh89} have given an 
asymptotic expression for $S^\omega_n$ for the case that $\omega$
is a fixed point of a primitive substitution and $\psi(\omega)$ 
depends only on $\omega_0$ (here $\Omega\subset\{0,\ldots,a\}^\ZZ$).
They find,  under certain assumptions on the eigenvalues of the 
substitution matrix, an upperbound
$\beta<1$ on $\gamma(\omega)$ if $\int\psi\,d\nu=0$.
This is a result for one particular sequence in $\Omega$, a fixed
point of the substitution, 
and therefore  has no direct implication for the H\"older continuity
of $\mu_\psi$ at 0.

\section{Twisted generalized random walk}

The exponential sums 
\beq\label{e4.1}
 \snol := \sum_{k=0}^{n-1} \char{\lambda k} \psi(T^k \omega)
\eeq
define a walk in the complex plane for each $\omega\in\Omega$ 
and each $\lambda\in\TT$.
One can think of these as walks with a rotational bias:
at step $k$ the walker changes its direction by $\char\lambda$
and then makes a step of length $|\psi(T^k\omega)|$ in the 
direction $\mbox{\rm arg}(\psi(T^k\omega))$. 
When plotted in the complex plane, such walks often give rise
to pretty pictures \cite{DeMe81,Aub+88}. 
The mean squared displacement $\int|\snol|^2\,d\nu(\omega)$ of $\snol$ is 
given by $nG_n(\lambda)$, and now depends on $\lambda$.
By Theorem~\ref{thm1}, its critical exponent $\gmsl$ satifies
$\gmsl\leq2-\alpha$ on any interval where $\mu_\psi$ is \UH\alpha.
Again, superdiffusive behavior of $\int|\snol|^2\,d\nu(\omega)$
requires singularity of $\mu_\psi$ at $\lambda$.

The aim of this section is to show that $\snol$ is itself
a GRW (i.e., a GRW for some ergodic dynamical system), except possibly
for a countable set of $\lambda$, and to relate the critical
exponents $\gamma(\lambda)$ of $|\snol|^2$ to $\mu_\psi$.
This will explain the meaning of the method of Aubry et al.\
\cite{Aub+87a,Aub+88}, see the next section.
It is of interest because it is often easier to deal numerically
with one---hopefully typical---trajectory then with 
$\int|\snol|^2\,d\nu(\omega)$.

An eigenvalue of $(\Omega, T, \nu)$ is a $\lambda\in[0,1] \simeq\TT$ for which
there is a $\phi\in\Ltwo(\Omega,\nu)$ such that $U\phi=\cchar\lambda \phi$.
The set of eigenvalues forms a countable group, which will be denoted by
$\Lambda$.
Let $\QQ[\Lambda]:=\bigl\{\mu\in\TT \bigm| \mu=q\lambda, 
\lambda\in\Lambda, q\in\QQ \bigr\}$. 

\begin{proposition}\label{prop1}
For $\lambda\not\in\QQ[\Lambda]$ there exists a 
$\gamma(\lambda)\in[0,\gmsl]$ such that 
$\gamma(\lambda, \omega) = \gamma(\lambda)$ for $\nu$-\ae $\omega$.
\end{proposition}

\begin{proof}
Let $X$ be a compact metric space and $\rho$ a Borel probability
measure on $X$ that is ergodic for an invertible, 
measurable transformation $R$ on $X$.
The direct product of $(\Omega,T,\nu)$ and $(X,R,\rho)$ is the
dynamical system $(\tilde\Omega,\tilde T,\tilde\nu)$ defined 
by $\tilde\Omega:= \Omega \times X$, $\tilde T(\omega, x):=
(T\omega, Rx)$, $\tilde\nu:= \mu\times \rho$.
This direct product is ergodic if and only if $(\Omega,T,\nu)$ 
and $(X,R,\rho)$ share no eigenvalues other than 0
(see e.g.\ \cite{CFS}, Theorem~10.1.1).

If $\lambda$ is irrational, take $X=\TT$, $\rho$ normalized
Lebesgue measure and let $R=R_\lambda$ be defined by
$R_\lambda x := x +\lambda$ (mod 1).
If $\lambda= p/q$ is rational, take $X=\{k p/q\}_{k=0}^{q-1}$, 
let $\rho$ be normalized counting measure on $X$ and again let $R=R_\lambda$.
In both cases $(X,R_\lambda,\rho)$ is ergodic and has eigenvalues
$\char{m\lambda}$, $m\in\ZZ$.
Therefore the direct product $(\tilde\Omega,\tilde T, \tilde\nu)$
is ergodic if $\lambda\not\in\QQ[\Lambda]$.
Take such a $\lambda$.

For $\psi\in\Ltwo(\Omega,\nu)$ let $\tilde\psi\in\Ltwo(\tilde\Omega,
\tilde\nu)$ be defined by $\tilde\psi(\omega, x) := \char{x} \psi(\omega)$.
Let $\tilde U$ act on $\Ltwo(\tilde\Omega, \tilde\nu)$ by
$\tilde U \phi = \phi \circ \tilde T$, and define  
$\tilde S_n^{(\omega, x)} := \sum_{k=0}^{n-1} 
\tilde U^k \tilde\psi (\omega, x)$.
(Note that $\tilde S_n^{(\omega, x)}$ implicitly depends on $\lambda$.)
Then $\snol = \tilde S_n^{(\omega, 0)}$.
Since $|\tilde S_n^{(\omega, x)}|$ is independent of $x$, the critical
exponent of $\int |\tilde S_n^{(\,\cdot\,)}|^2\,d\tilde\nu$ is the
same as the critical exponent $\gmsl$ of $\int |\snol|^2\,d\nu(\omega)$.
By Lemma~\ref{lem1} there is a $\gamma(\lambda)\in[0,\gmsl]$
such that the critical exponent $\tilde\gamma(\lambda, \omega, x)$
is equal to $\gamma(\lambda)$ for $\tilde\nu$-\ae $(\omega, x)$.
But $\tilde\gamma(\lambda,\omega, x)=\tilde\gamma(\lambda, \omega, 0)
=\gamma(\lambda,\omega)$ for all $x\in X$, so 
$\gamma(\lambda,\omega)=\gamma(\lambda)$ for $\nu$-\ae $\omega$.
\end{proof}

\begin{corollary}\label{cor1}
If $(\Omega, T, \nu)$ is weakly mixing then there exists for 
all $\lambda\in\TT$ a $\gamma(\lambda)\in[0,\gmsl]$ such that
$\gamma(\lambda,\omega)=\gamma(\lambda)$ for $\nu$-\ae $\omega$.
\end{corollary}
\begin{proof}
Recall that, by definition, $(\Omega,T,\nu)$ is weakly mixing
if it has no eigenvalues other than 1.
The case $\lambda=0$ is covered by Lemma~\ref{lem1}.
\end{proof}

\begin{corollary}\label{cor2}
Suppose $\lambda\not\in\QQ[\Lambda]$ and $\gamma(\lambda)>1$.
Let $0<\epsilon < \gamma(\lambda)-1$.
Then $\mu_\psi$ is not \UH{(2-\gamma(\lambda)+\epsilon)} on
any neighborhood of $\lambda$.
\end{corollary}
\begin{proof}
Let $\beta:=\gamma(\lambda)-\epsilon$.
Then $1<\beta<\gmsl$ and 
$\limsup_{n\to\infty} n^{-(\beta-1)}G_n(\lambda) =\infty$.
Let $\Delta$ be any neighborhood of $\lambda$.
By Theorem~\ref{thm1}, $\mu_\psi$ is not \UH{(2-\beta)} on $\Delta$.
\end{proof}

Note that the exceptional set in Proposition~\ref{prop1} depends on $\lambda$.

Corollary~\ref{cor2} shows that a single (but typical) trajectory 
provides upper bounds on the degree of H\"older continuity of
$\mu_\psi$.
As explained in Remark~\ref{scpart}, a degree of H\"older
continuity strictly less than one does not prove that a 
measure is purely singular continuous.
Thus the question arises whether one can determine from 
a single (typical) trajectory that the spectral measure
has no absolutely continuous part.
In general, this is not possible, as is explained below.

The following lemma is well known.

\begin{lemma}\label{lem2}
$\lim_{n\to\infty} n^{-1} \bigl|\snol|^2 \, d\lambda = \mu_\psi$
weakly for $\nu$-\ae $\omega$.
If $(\Omega,T,\nu)$ is uniquely ergodic and $\psi$ is continuous then 
`$\nu$-\ae $\omega$' can be replaced by `for all $\omega\in\Omega$'.
\end{lemma}
\begin{proof}
For $p\in\ZZ$,
\beq\label{e4.2}
 \int \char{\lambda p} \snol \,d\lambda 
   = \frac 1 n \sum_{m=0}^{n-1} \psi( T^{m-p}\omega ) 
       \overline{\psi( T^m \omega )},
\eeq
which by the ergodic theorem converges to $(U^{-p} \psi,\psi)$ as 
$n\to\infty$ for $\nu$-\ae $\omega$. 
Hence the Fourier coefficients of $\snol\,d\lambda$ converge
pointwise to those of $\mu_\psi$, for $\nu$-\ae $\omega$.
This shows that 
$\lim_{n\to\infty} n^{-1} \bigl|\snol|^2 \, d\lambda = \mu_\psi$ weakly
for $\nu$-\ae $\omega$.
The averages (\ref{e4.2}) converge for all $\omega\in\Omega$
as $n\to\infty$ if $(\Omega, T, \nu)$ is uniquely ergodic
and $\psi$ is continuous (e.g. \cite{CFS}, Theorem~1.8.2).
\end{proof}

Thus both $\frac1n |\snol|^2\, d\lambda$ (for a.e.-$\omega$)
and its expectation $G_n(\lambda)\, d\lambda$ converge
weakly to $\mu_\psi$ as $n\to\infty$.
There is a difference, however.
The sequence $G_n(\lambda)$ converges for Lebesgue-\ae
$\lambda$ and its limit is the value of the density of the 
absolutely continuous part of $\mu_\psi$ (see
Remark~\ref{beta01}).
But $\lim \frac 1n |\snol|^2$ in general does {\em not}
exist for Lebesgue-\ae $\lambda$.
(E.g., if $\psi(T^k \omega)$, $k\in\ZZ$,
are independent and uniformly distributed in $\TT$ then for every 
$\lambda$ and $\nu$-\ae $\omega$ the limit
does not exist.)

Now assume that $(\Omega, T, \nu)$ is uniquely ergodic and
that $\psi$ is continuous.
Then, if $\lim_{n\to\infty} \frac 1n |\snol|^2$ exists for
Lebesgue-\ae $\lambda$ in some interval,
the limit is equal to the density of $\mu_\psi$ with respect
to the Lebesgue measure in that interval.
In particular, if $\lim_{n\to\infty} \frac 1n |\snol|^2=0$ \ae on $\TT$,
then $\mu_\psi$ has no absolutely continuous part.
In this special case examining a single trajectory allows
to conclude that the spectral measure has no absolutely
continous part.

\smallskip
\remarks\label{rem4.1}
The results in Sections~3 and 4 generalize to
flows, if, for $T>0$ and $\lambda\in\RR$, $S_T^\omega(\lambda)$ is 
defined by $S_T^\omega(\lambda) := \int_0^T \char{\lambda x} 
\psi(T_t \omega)\, dt$.
An eigenvalue of the flow $T_t$ is a $\lambda\in\RR$ for which
there is a $\phi\in\Ltwo(\Omega, \nu)$ such that 
$U_t \phi = \cchar{\lambda t} \phi$.

\nextrem
If $\mu_\psi(\{\lambda\})>0$ then $S^\omega_n(\lambda)$
behaves ballistically, $\gmsl=2$, with diffusion coefficient 
$\mu_\psi(\{\lambda\})$, see Remark~\ref{beta01}.
If $(\Omega,T,\nu)$ is uniquely
ergodic and the eigenfunction $\phi$ for $\lambda$ is continuous
then $\gamma(\lambda,\omega)=2$ for all $\omega\in\Omega$ \cite{Rob94}.

\section{Applications}

\subsection{The model of Aubry et al.}

Aubry et al.\ \cite{Aub+87a,Aub+88} have considered a model of
atoms on the line defined by
\beq\label{e5.1}
  x_n - x_{n-1} = 1 + \xi 1_{[0,\beta)}( n\alpha+\theta ),
\eeq
where $\{x_n\}$ is the set of atomic positions, $\alpha$
is irrational, and $x_0\in\RR$, $\beta, \theta \in\TT$ and $\xi >0$ 
are parameters. 
(Changing $x_0$ translates the structure; $\theta$ selects 
a structure within a `local isomorphism class' determined 
by $\alpha$, $\beta$ and $\xi$.)
They were interested in the `structure factor' $S(\lambda)$, 
an unbounded measure  defined by
\beq\label{e5.2}
  dS(\lambda) = \lim_{T\to\infty} T^{-1} 
     \Bigl| \sum_{x_n\in[0,T]} \char{ \lambda x_n} \Bigr|^2\, d\lambda,
\eeq
where the limit is taken in the vague topology (cf.\ \cite{Hof95b,Hof95pe}).
The structure factor does not depend on the choice of $x_0$ or $\theta$.
Aubry et al.\ argued that $S$ should be purely singular continuous 
if $\beta \not= n\alpha$ (mod 1).
One reason for this assertion was that they found critical
exponents $\eta(\lambda)$ of the sums 
\beq\label{e5.3}
  F_T(\lambda) :=  \sum_{x_n\in[0,T]} \char{ \lambda x_n} 
\eeq
to be strictly between $1/2$ and 1.
For the case $\beta=1/2$, $\alpha= \tau^{-1}$ ($\tau=(\sqrt 5 + 1)/2$)
they showed analytically \cite{Aub+88} that $\eta(\lambda) = c$
for some $\frac12<c<1$ and $\lambda$ in a countable dense set.

This section explains that $2-2\eta(\lambda)$ is an upperbound
on the degree of H\"older continuity of $S$ on neighborhoods of $\lambda$.
The idea is to view $S$ as a limit of spectral measures of a dynamical
system, the flow under the function $f(x)= 1 + \xi 1_{[0,\beta)}(x)$
over the irrational rotation $\alpha$ on $\TT$.
This was explained in detail in \cite{Hof96a}.
The essence is the following.
Let $\phi\geq 0$ be a $C^\infty$-function with support in $[-1/2, 1/2]$,
$\int \phi(x)\,dx=1$, and for $1>\epsilon>0$ let 
$\phi_\epsilon(x) := \epsilon^{-1} \phi(x/ \epsilon)$.
Then the function $\rho_\epsilon := \phi_\epsilon \ast \sum_{n\in\ZZ}
\delta_{x_n}$ is of the form $x\to \psi_\epsilon (T_x \omega)$
for an $\omega\in\Omega$ and a $\psi_\epsilon\in\Ltwo(\Omega, \nu)$,
where $\Omega=\Omega_{\beta, \xi} =
 \bigl\{ (x,y) \bigm| x\in\TT,\;\; 0\leq y < f(x) \bigr\}$
and $\nu$ is the normalized Lebesgue measure on $\Omega$.
The spectral measure $\mu_{\psi_\epsilon}$ of $\psi_\epsilon$ satisfies
\beq\label{e5.4} 
  \mu_{\psi_\epsilon} = S |\hat\phi_\epsilon|^2,
\eeq
where $\hat\phi$ denotes the Fourier transform of $\phi$.
Thus $\mu_{\psi_\epsilon} \to S$ vaguely as $\epsilon\to 0$.
Since $\hat\phi_\epsilon$ is smooth is follows that for every interval
$I$ there is an $\epsilon$ such that the degrees of H\"older
continuity of $S$ and $\mu_{\psi_\epsilon}$ are the same on $I$.

Now consider the integral $S_T^\omega(\lambda):=
\int_0^T \char {\lambda x} \psi_\epsilon(T_x\omega)\, dx$.
For each $\omega\in\Omega$ there is an $A_\omega$, $|A_\omega|\leq 2$
such that 
\begin{eqnarray*}
 S^\omega_T (\lambda) 
  &=& 
   \int_0^\infty \char{\lambda x }
    \Bigl(\sum_{x_n\in[0,T]} \phi_\epsilon \ast \delta_{x_n} \Bigr)
      (x) \, dx + A_\omega \\ 
  &=& \hat\phi_\epsilon(\lambda) F_T(\lambda) + A_\omega
\end{eqnarray*}
It follows that the critical exponent $\eta(\lambda)$ of 
$F_L(\lambda)$ is related to the critical exponent $\gamma(\lambda)$
of $|S_L^\omega (\lambda)|^2$ by $\gamma(\lambda)= 2\eta(\lambda)$.
Hence corollary~\ref{cor2} (in `flow form', cf.\ Remark~\ref{rem4.1})
gives that $S$ is not \UH{(2-2\eta(\lambda) + \delta)} on any
neighborhood of $\lambda$, for all $0<\delta<2\eta(\lambda)-1$.

\smallskip
\remark
The connection with the flow under the function $f$ was used in
\cite{Hof96a} to determine the parameters $\alpha$, $\beta$, 
$\xi$ for which the unbounded measure $S$ is continuous 
(apart from a delta function at zero, which is present for
all parameters).
This in turn was used to prove that
for every irrational $\xi$ and every $\beta$ the 
measure $S$ is purely singular continuous for generic $\alpha$
(i.e., for $\alpha$ in a dense $G_\delta$).
We recently found a paper by Goodson and Whitman \cite{GoWh80}
that proves by periodic approximation that for each $\xi$
(irrational or not) and each irrational $\alpha$ the flow under
$f$ has singular spectrum (i.e., no spectral measure has an 
absolutely continuous part) for Lebesgue-\ae $\beta$.
Combined with the continuity result of \cite{Hof96a}
this gives that for every irrational $\xi$ and
every irrational $\alpha$, the flow has purely singular
continuous spectrum for almost every $\beta$.
For these parameters $S$ is purely singular continuous
apart from the delta function a zero.
For $\beta=1/2$, $\alpha=\tau^{-1}$, the parameters considered
in \cite{Aub+88}, $S$ is continuous \cite{Hof96a}.
It has not been proved that it is singular continuous
although the work of Aubry et al.\ does suggest that it is 
singular continuous.

\subsection{Energy growth in driven oscillators}

Bunimovitch et al.\ \cite{Bun+91} consider a classical driven
oscillator with natural frequency $\omega_0$ described by the 
Hamiltonian $H=p^2/2 +\omega_0 q^2/2 - q \psi (T_t x)$, where 
$T_t$ is an ergodic flow on a compact metric space 
with invariant probability measure $\nu$. 
They are interested in the expectation of
the time evolution $\langle E_t\rangle_\nu$ of the energy 
of the oscillator. 
Their equations (2.9) and (2.10) can be rewritten to give
\[
  \langle E_t \rangle_\nu= \omega_0^2 q_0^2/2 + p_0^2/2 
        + {\omega_0^2 \over 2}\, t G_t (\omega_0)
\]
where $G_t$ is as in Remark~\ref{rem2.3}.
Bunimovich et al.\ express this in terms of the correlation
function $\hat\mu_\psi$.
In Proposition~3.1 they show that $\langle E_t\rangle_\nu$
grows linearly with $t$ and has a well defined diffusion 
coefficient if the correlation functions decays
sufficiently fast. 
Remark~\ref{diff} generalizes this: it suffices to require that
$\mu_\psi$ is absolutely continuous on a neighborhood of 
of $\omega_0$ with  a density that can be chosen to be continuous 
at $\omega_0$.
The correlation function need not decay at all in order to have
diffusive behavior of $\langle E_t \rangle_\nu$.

Again, superdiffusive behavior of $\langle E_t\rangle$ requires
$\mu_\psi$ to be singular at $\omega_0$, in the sense that 
$\mu_\psi$ is not \UH\alpha\ on any neighborhood of $\omega_0$,
for some $\alpha<1$.

Note that by Proposition~\ref{prop1} the critical exponent of $E_t$
is independent of $x$ ($\nu$-a.e.), for all but possibly 
countably many $\omega_0$.

The expection of the energy growth for driven quantum oscillators
is also determined by $tG_t$ \cite{Bun+91,Com92}.

\section*{Acknowledgements}
It is a pleasure to thank F.M.~Dekking and O.~Knill for
discussions.


\begin{thebibliography}{10}

\bibitem{AaKe82}
J.~Aaronson and M.~Keane.
\newblock The visits to zero of some deterministic random walks.
\newblock {\em Proc. London Math. Soc. (3)}, 44:535--553, 1982.

\bibitem{Aub+87a}
S.~Aubry, C.~Godr\`eche, and J.M. Luck.
\newblock A structure intermediate between quasi-periodic and random.
\newblock {\em Europhys. Lett.}, 4:639--643, 1987.

\bibitem{Aub+88}
S.~Aubry, C.~Godr\`eche, and J.M. Luck.
\newblock Scaling properties of a structure intermediate between quasiperiodic
  and random.
\newblock {\em J. Stat. Phys.}, 51:1033--1074, 1988.

\bibitem{Bar+96}
J.M. Barbaroux, J.M. Combes, and R.~Montchio.
\newblock Remarks on the relation between quantum dynamics and fractal spectra.
\newblock preprint Marseille CPT-96/P.3303.

\bibitem{Ber81}
H.~Berbee.
\newblock Recurrence and transience for random walks with stationary
  increments.
\newblock {\em Z. Wahr. Verw. Geb.}, 56:531--536, 1981.

\bibitem{Bun+91}
L.~Bunimovich, H.R. Jauslin, J.L. Lebowitz~A. Pellegrenotti, and P.~Nielaba.
\newblock Diffusive energy growth in classical and quantum driven oscillators.
\newblock {\em J. Stat. Phys.}, 62:793--817, 1991.

\bibitem{Com93}
J.M. Combes.
\newblock Connections between quantum dynamics and spectral properties of time
  evolution operators.
\newblock In W.F. Ames, E.M Harrell, and J.V. Herod, editors, {\em Differential
  Equations with Applications to Mathematical Physics}. Academic Press, Inc.,
  1993.

\bibitem{Com92}
M.~Combescure.
\newblock Recurrent versus diffusive dynamics for a kicked quantum oscillator.
\newblock {\em Ann. Inst. Henri Poincar\'e Phys. Th\'eor.}, 57:67--87, 1992.

\bibitem{CFS}
I.P. Cornfeld, S.V. Fomin, and Ya.G. Sinai.
\newblock {\em Ergodic Theory}, volume 115 of {\em {Grundlehren} der
  mathematischen {Wissenschaften} in {Einzeldarstellungen}}.
\newblock Springer Verlag, 1982.

\bibitem{Oli95}
C.R. de~Oliveira.
\newblock Numerical study of the long-time behaviour of quantum systems driven
  by {Thue-Morse} forces. {A}pplication to two-level systems.
\newblock {\em Europhys. Lett.}, 31:63--68, 1995.

\bibitem{Dek82}
F.M. Dekking.
\newblock On transience and recurrence of generalized random walks.
\newblock {\em Z. Wahr. Verw. Geb.}, 61:459--465, 1982.

\bibitem{Dek93}
F.M. Dekking.
\newblock Marches automatiques.
\newblock {\em J. de Th\'eorie de Nombres de Bordeaux}, 5:93--100, 1993.

\bibitem{Dek95}
F.M. Dekking.
\newblock Random and automatic walks.
\newblock In F.~Axel and D.~Gratias, editors, {\em Beyond Quasicrystals}. Les
  \'Editions de Physique/Springer, 1995.

\bibitem{DeMe81}
F.M. Dekking and M.~Mend\`es France.
\newblock Uniform distribution modulo one: A geometric point of view.
\newblock {\em J. Reine Angew. Math.}, 329:143--153, 1981.

\bibitem{Rio+96}
R.~del Rio, S.~Jitomirskaya, Y.~Last, and B.~Simon.
\newblock Operators with singular continuous spectrum, {IV}. {Hausdorff}
  dimensions, rank one perturbations, and localization.
\newblock {\em J. Analyse Math.}, To appear.

\bibitem{DuTh89}
J.M. Dumont and A.~Thomas.
\newblock Syst\`emes de num\'eration et fonctions fractales relatifs aux
  substitutions.
\newblock {\em Theor. Comp. Science}, 65:153--169, 1989.

\bibitem{GoWh80}
G.R. Goodson and P.N. Whitman.
\newblock On the spectral properties of a class of special flows.
\newblock {\em J. Lond. Math. Soc.}, 21:567--576, 1980.

\bibitem{Gua89}
I.~Guarneri.
\newblock Spectral properties of quantum diffusion on discrete lattices.
\newblock {\em Europhys. Lett.}, 10:95--100, 1989.

\bibitem{Gua93}
I.~Guarneri.
\newblock On an estimate concerning quantum diffusion in the presence of a
  fractal spectrum.
\newblock {\em Europhys. Lett.}, 21:729--733, 1993.

\bibitem{GuMa94}
I.~Guarneri and G.~Mantica.
\newblock On the asymptotic properties of quantum dynamics in the presence of a
  fractal spectrum.
\newblock {\em Ann. Inst. Henri Poincar\'e Phys. Th\'eor.}, 61:369--379, 1994.

\bibitem{Hof95b}
A.~Hof.
\newblock On diffraction by aperiodic structures.
\newblock {\em Commun. Math. Phys.}, 169:25--43, 1995.

\bibitem{Hof96a}
A.~Hof.
\newblock On a `{S}tructure intermediate between quasiperiodic and random'.
\newblock {\em J. Stat. Phys.}, 84:309--320, 1996.

\bibitem{Hof95pe}
A.~Hof.
\newblock Diffraction by aperiodic structures.
\newblock In R.V. Moody and J.~Patera, editors, {\em Proceedings of the NATO
  Advanced Study Institute on the Mathematics of Aperiodic Long Range Order}.
  Kluwer Academic Publishers, To appear.

\bibitem{JiLa96}
S.~Jitomirskaya and Y.~Last.
\newblock Dimensional {Hausdorff} properties of singular continuous spectra.
\newblock {\em Phys. Rev. Lett.}, 76:1765--1769, 1996.

\bibitem{Katznelson}
Y.~Katznelson.
\newblock {\em An Introduction to Harmonic Analysis.}
\newblock John Wiley \& Sons, 1968.

\bibitem{Las96}
Y.~Last.
\newblock Quantum dynamics and decompositions of singular continuous spectra.
\newblock {\em J. Func. Anal.}, To appear.

\bibitem{Lia87}
P.~Liardet.
\newblock Regularities of distribution.
\newblock {\em Comp. Math.}, 61:267--293, 1987.

\bibitem{Luc89}
J.M. Luck.
\newblock {Cantor} spectra and scaling of gap widths in deterministic aperiodic
  systems.
\newblock {\em Phys. Rev. B}, 39:5834--5849, 1989.

\bibitem{Luc+88}
J.M. Luck, H.~Orland, and U.~Smilansky.
\newblock On the response of a two-level quantum system to a class of
  time-dependent quasiperiodic perturbations.
\newblock {\em J. Stat. Phys.}, 53:551--564, 1988.

\bibitem{PiFe94}
A.S. Pikovsky and U.~Feudel.
\newblock Correlations and spectra of strange non-chaotic attractors.
\newblock {\em J. Phys. A: Math. Gen.}, 27:5209--5219, 1994.

\bibitem{Pik+95}
A.S. Pikovsky, M.A. Zaks, U.~Feudel, and J.~Kurths.
\newblock Singular continuous spectra in dissipative dynamics.
\newblock {\em Phys. Rev. B}, 52:285--296, 1995.

\bibitem{Queffelec}
M.~Queff\'elec.
\newblock {\em Substitution Dynamical Systems---Spectral Analysis}, volume 1294
  of {\em Lecture Notes in Mathematics}.
\newblock Springer, 1987.

\bibitem{Rob94}
E.A. {Robinson, Jr.}
\newblock On uniform convergence in the {Wiener}-{Wintner} theorem.
\newblock {\em J. Lond. Math. Soc.}, 49:493--501, 1994.

\bibitem{Rogers}
C.A. Rogers.
\newblock {\em Hausdorff measures}.
\newblock Cambridge University Press, 1970.

\bibitem{Str90}
R.S. Strichartz.
\newblock Fourier asymptotics of fractal measures.
\newblock {\em J. Func. Anal.}, 89:154--187, 1990.

\bibitem{WeWe92}
Z.-X. Wen and Z.-Y. Wen.
\newblock Marches sur les arbres homog\`enes suivant une suite substitutive.
\newblock {\em S\'em. T\'eor. Nombres, Bordeaux}, 4:155--186, 1992.

\bibitem{Zygmund}
A.~Zygmund.
\newblock {\em Trigonometric Series}.
\newblock Cambridge University Press, 1959.

\end{thebibliography}
\end{document}